# Functional Hypergraph Uncovers Novel Covariant Structures over Neurodevelopment

Shi Gu[a,d,e], Muzhi Yang[a,b,e], John D. Medaglia[c], Ruben C. Gur[a], Raquel E. Gur[a], Theodore D. Satterthwaite[a], Danielle S. Bassett[d,e,*]

[a]Department of Psychiatry, University of Pennsylvania, Philadelphia, PA 19104, USA
[b]Applied Mathematics and Computational Science Graduate Group, University of Pennsylvania, Philadelphia, PA 19104, USA
[c]Moss Rehabilitation Research Institute, Elkins Park, PA 19027
[d]Department of Electrical and Systems Engineering, University of Pennsylvania, Philadelphia, PA 19104, USA
[e]Department of Bioengineering, University of Pennsylvania, Philadelphia, PA

**Corresponding Author:**

Danielle S. Bassett
Office: 113 Hayden Hall
T: 215-746-1754
F: 215-573-2071
Email: dsb@seas.upenn.edu
Mail: 210 S. 33rd Street
240 Skirkanich Hall
Philadelphia, PA 19104-6321

## Abstract

Brain development during adolescence is marked by substantial changes in brain structure and function, leading to a stable network topology in adulthood. However, most prior work has examined the data through the lens of brain areas connected to one another in large-scale functional networks. Here, we apply a recently-developed hypergraph approach that treats network connections (edges) rather than brain regions as the unit of interest, allowing us to describe functional network topology from a fundamentally different perspective. Capitalizing on a sample of 780 youth imaged as part of the Philadelphia Neurodevelopmental Cohort, this hypergraph representation of resting-state functional MRI data reveals three distinct classes of sub-networks (hyperedges): clusters, bridges, and stars, which represent spatially distributed, bipartite, and focal architectures, respectively. Cluster hyperedges show a strong resemblance to the functional modules of the brain including somatomotor, visual, default mode, and salience systems. In contrast, star hyperedges represent highly localized subnetworks centered on a small set of regions, and are distributed across the entire cortex. Finally, bridge hyperedges link clusters and stars in a core-periphery organization. Notably, developmental changes within hyperedges are ordered in a similar core-periphery fashion, with the greatest developmental effects occurring in networked hyperedges within the functional core. Taken together, these results emphasize that the network organization of human brain emerges across multiple scales and evolves substantially through the adolescent period.

## Keywords:

brain network, core-periphery, functional connectivity, network modules

# Introduction

Understanding the human brain as a networked system has offered important insights into the development of brain function across the lifespan (Bassett and Sporns 2016). In this perspective, brain regions are treated as network nodes, and functional connections between brain regions are treated as network edges (Bullmore and Bassett 2011). Studies examining functional brain networks at rest measured with functional MRI have revealed a modular functional organization (Dosenbach et al. 2007; Power et al. 2011) comprised of reproducible network communities such as the default-mode (Raichle et al. 2001), cognitive control (Sridharan et al. 2008), visual (Lowe et al. 1998), and somatomotor (Biswal et al. 1995) systems. These modules evolve considerably during development in adolescence (Power et al. 2010; Satterthwaite, Wolf, et al. 2013; Gu et al. 2015), and are thought to allow for the well-known expansion of cognitive and behavioral capabilities that defines this period.

While such studies have offered critical insights into the network neurophysiology of development, they constitute relatively coarse levels of interrogation. Modules are mesoscale structures, defined as sets of brain areas. As the module structure depends on the average properties of network edges, they are relatively insensitive to how these edges are combined with each other. In order to resolve this level of detail, we take an alternative approach by treating the network edge as the unit of interest (Bassett et al. 2014; Davison et al. 2015, 2016). This choice is guided by the fact that edges may develop differentially in a coordinated fashion over the lifespan (Davison et al. 2016), leading to architectural features that cannot simply be characterized by modules or the nodes that compose them (Bassett et al. 2014). Intuitively, such developmental coordination of functional connections may be driven by intrinsic computations

Functional Hypergraph

(Bassett et al. 2014), and subsequently have mutually trophic effects on underlying structural connectivity (Bassett et al. 2008). From a computational standpoint, co-modulated functional connections can be thought of as circuits -- edges that link disparate computational units -- that may form more fundamental structures that prefigure the emergence of well-described cognitive systems observed in adulthood.

To investigate functional brain network architecture at this finer scale, we examine high-resolution edge-based hypergraphs in a large sample of youth imaged as part of the Philadelphia Neurodevelopmental Cohort. In contrast to typical functional networks where nodes represent brain regions, hypergraphs are built on the pairwise correlations between network edges across individuals, allowing for the detection of groups of coherent edges known as hyperedges. This particular emphasis on functional edges enables us to address several specific hypotheses. First, we expected that hypergraphs would corroborate and extend findings from prior (region-based) network analyses, and reveal cluster hyperedges that form densely interconnected brain systems. Second, we hypothesized that this approach would allow us to uncover novel types of subgraphs that were distinguishable from traditional network modules. Third and finally, we hypothesized that the evolution of connectivity during adolescence would differ by hyperedge type. As described below, such edge-based analyses allowed us to uncover novel fine-scaled functional subgraphs that display differential patterns of development during adolescence.

## Materials and Methods

**Data acquisition and preprocessing**

Data were acquired in a collaboration between the Center for Applied Genomics (CAG) at Children's Hospital of Philadelphia (CHOP) and the Brain Behavior Laboratory at the University of Pennsylvania (Penn). Study procedures were reviewed and approved by the Institutional Review Board of both CHOP and Penn. Resting state fMRI data were acquired from 780 healthy youth ages 8 - 22 years (Satterthwaite, Elliott, et al. 2014).

All subject data were acquired on the same scanner (Siemens Tim Trio 3 Tesla, Erlangen, Germany; 32 channel head coil) using the same imaging sequences. Blood oxygen level dependent (BOLD) fMRI was acquired using a whole-brain, single-shot, multi-slice, gradient-echo (GE) echoplanar (EPI) sequence of 124 volumes with the following parameters: TR/TE=3000/32 ms, flip=90 degrees, FOV=192 $\times$ 192 mm, matrix=64 $\times$ 64, slice thickness/gap =3mm/0mm. The resulting nominal voxel size was 3.0 $\times$ 3.0 $\times$ 3.0 mm. A fixation cross was displayed as images were acquired. Subjects were instructed to stay awake, keep their eyes open, fixate on the displayed crosshair, and remain still.

Functional imaging used tools from FSL (FMRIB's Software Library) and AFNI with a preprocessing scheme described elsewhere (Satterthwaite, Elliott, et al. 2013; Satterthwaite, Wolf, et al. 2014). Briefly, time-series were processed with a confound regression technique optimized to reduce the influence of subject motion. Specifically, the first 4 volumes were removed for initial signal stabilization, resulting in 120 remaining volumes for subsequent analyses. fMRI time series were band-pass filtered to retain frequencies between 0.01 and 0.08

Hz. Functional images were re-aligned using MCFLIRT. BET was used to remove non-brain areas from structural images via skull-stripping (Smith 2002). Further details about the preprocessing can be found in the SI. The data reported in this paper have been deposited in the database of Genotypes and Phenotypes (dbGaP), www.ncbi.nlm.nih.gov/gap (accession no. phs000607.v1.p1).

**Functional network construction**

We extracted regional mean BOLD time series from 264 functionally defined regions covering cortical and subcortical areas (Power et al. 2011). Consistent with prior work (Bassett, Wymbs, et al. 2011, 2013), we estimated functional connectivity $A_{ij}$ between any two pairs of regions i and j using a wavelet coherence (Grinsted et al. 2004) in the frequency interval approximately 0.01--0.08 Hz. The fully weighted adjacency matrix **A** therefore represents the functional brain network for a given subject in which N network nodes represent brain regions and $E$ network edges represent functional connections between those regions (Zhang et al. 2016).

**Hypergraph construction**

To construct an edge-by-edge hypergraph (Berge and Minieka 1973), we stacked subject adjacency matrices to create a 3-dimensional adjacency tensor with elements $A_{ijs}$, where s indexes over subjects (Bassett et al. 2014). For an edge connecting a given pair of regions i and j, the elements $A_{ijs}$, for all s could be treated as a vector: a time series of observations over the entire sample (N=780). For every pair of edges, $A_{ijs}$, and $A_{kls}$ for all s, we computed the Pearson correlation coefficient between these time series $H_{mn}$, where m indexes over edge pairs i and j and n indexes over edge pairs k and l, and we stored these values in the E × E hypergraph **H**.

Following (Bassett et al. 2014) and to control the false positive rate, we thresholded the matrix **H** by setting all correlation coefficient values $H_{mn}$ to zero whose respective p-values were greater than 0.05. Intuitively, the hypergraph **H** provides a cross-sectional representation of functional connections that co-vary with one another. Entries are positive if the weights of the corresponding edges are positively correlated over subjects, and entries are negative if the weights of the corresponding edges are negatively correlated over subjects.

**Hyperedge archetypes**

We identified hyperedges -- significant clusters of co-varying edges -- by applying a common network-based community detection algorithm (Porter et al. 2009; Fortunato 2010) to the hypergraph **H**. Next, we defined hyperedge archetypes that each displayed interpretable spatial configurations in the brain, and that corresponded to differential dynamic processes. For each hyperedge, we listed the edges that composed the associated cluster r, and then determined the set of nodes (brain regions) that were touched by at least one of those edges. We then defined a binary adjacency matrix $\mathbf{B}^r$ whose elements $B_{ij}^r$ indicated the presence (1) or absence (0) of an edge between nodes i and j that were also present in the cluster r. We observed that the matrices $\mathbf{B}^r$ fell into one of three categories: focal star hyperedge, bridging bipartite hyperedge, and clustering hyperedge (see Figure 2). Stars consist of edges that are linked to one another via one, two, or three nodes. Bridges consist of edges that connect one set of nodes to a second set of nodes, but do not connect nodes within the same set. Clusters consist of edges that connect nodes both within and between sets. See SI for a detailed description of the clustering method and Suppl. Figure. S3 for details on parameter choices.

**Hyperedge connector estimation**

Bridges are particularly interesting as they represent the collection of edges that can connect two or more hyperedges with one another. We hypothesized that the bridges connected clusters to stars. To test our hypothesis, we performed the following analysis: First, we set the valid stars and the clusters as fundamental modules. Second, for each valid bridge, we computed the number of overlap regions on each side with every fundamental module and calculate the p-values versus the null distribution where each side of the bridge is randomly chosen from among all possible regions while preserving size. Third, we thresholded the p-values in Step 2 by applying a Bonferroni procedure to control the family-wise error rate (FWER) less than 0.05. Finally, according to the type of module on each end of the bridge, we classified the bridge into three sub-types of connectors: star-star connector, star-cluster connector, and cluster-cluster connector, and we counted how many passed tests in each sub-type divided by the corresponding numbers of possible tests for the normalization. Here the family-wise error rate (FWER) is used with Bonferroni's procedure for multiple comparisons correction.

**Edge correlation comparison**

For the edge comparison in Figure 3, we computed the correlation among edges within each predefined module and compared its distribution with that of the average correlation of edge pairs within the stars centered within the same module. FDR was corrected via Storey's approach (Storey 2002).

**Linear model of developmental effects**

Functional Hypergraph

To test which cluster hyperedges displayed an increasing strength with age, we mapped back the 6 uncovered clusters to each subject, computed the average strength of the masked connections and investigated its linear dependence on age and movement.

See supplement for additional methodological details.

**Results**

We constructed high-resolution hypergraphs (Figure 1) using resting state fMRI data acquired from 780 youth between the ages of 8 and 22 years (Satterthwaite, Elliott, et al. 2014). For each participant, we created adjacency matrices by calculating the wavelet coherence between all 34,716 pairs of 264 functionally-defined regions. Adjacency matrices were concatenated across subjects to create a three-dimensional matrix, and then collapsed to an edge-by-edge matrix whose elements were given by the Pearson correlation coefficient between each pair of edges. This hypergraph of dimensions $34,716 \times 34,716$ summarized the degree to which functional connections co-varied with one another over subjects.

**Hyperedges reveal novel architectural motifs**

Within the full hypergraph, we identified statistically significant functional hyperedges: groups of edges that co-varied in strength over subjects (see Methods) (Bassett et al. 2014). In the 363 significant hyperedges (Figure 1E), we algorithmically detect 3 distinct topological classes (Figure 2): *stars* (326 of 363), *bridges* (31 of 363), and *clusters* (6 of 363). Stars were the most numerous hyperedges in the hyergraph. In mathematics, a star graph is one in which edges emanate from a small set of nodes (Figure 2A). These star-like structures indicate the presence of neurophysiological drivers of functional connectivity that are localized to very few ($\leq 3$) brain regions. Bridges are bipartite graphs that are composed of edges connecting two separate sets of nodes (Figure 2B). These bridging structures suggest the existence of connectors among hyperedges. Clusters include edges that densely link spatially distributed regions. These hyperedges indicate the presence of coherent edge sets that are significantly co-modulated in their functional connectivity strength over subjects.

**Stars imply more cohesive collections of edges than predefined cognitive systems**

Stars were broadly distributed across the brain (Suppl. Figure. S2) unlike the well-known cognitive systems identified in traditional resting state analyses. To quantify this observation, we examined a set of systems or modules defined *a priori* (Power et al. 2011) and tested the null hypothesis that the cross-subject Pearson's correlation of edge weights of stars centered in nodes within a module was higher than edge weights within the module overall. This would establish whether stars are fundamentally more cohesive subunits than previously detected major systems. This approach revealed that stars were more coherent than all cognitive systems (FDR-corrected for multiple comparisions, $q < 0.001$), with the exception of the cerebellum (composed of 4 nodes) and memory retrieval systems (composed of 5 nodes). This result underscores the utility of the hypergraph approach in uncovering more coherent sub-structures than traditional community detections techniques uncovering network modules. It also suggests that star-shaped hyperedges may constitute one of the fundamental units of the brain's functional architecture.

**The functional core of cluster hyperedges**

We next turned to examining the nature of the cluster hyperedges, which occupy a central role in the hypergraph architecture. We observed that (Figure 4) clusters are remarkably similar to known functional sub-networks (Power et al. 2011). Of the six clusters identified, two were predominantly composed of regions in the default mode, two were predominantly composed of regions in visual cortex, one was largely composed of areas in somatosensory cortex, and one was largely composed of areas in the salience and cingular-opercular task control systems. These results demonstrate that cluster hyperedges re-capitulate previously-described large scale functional networks that have strongly coherent, dense connections. Also, see Suppl. Figure S2 for the spatial distribution of the cluster hyperedges.

**Bridges connect core clusters and peripheral stars**

Intuitively, bridges are groups of edges that can facilitate network integration by linking two distinct sets of brain areas (Figure 5A). Given that hyperedges neatly composed 3 distinct categories, we hypothesized that bridge hyperedges served to link the densely connected core of cluster hyperedges and less-connected star hyperedges. More specifically, the null hypothesis here was that bridges randomly connected two parts of the brain and were not significantly enriched for any of the following: *star-star* connections, *cluster-cluster* connections, and *star-cluster* connections (see *Methods*). We observed that bridges were far more likely to connect core cluster hypergraphs to peripheral stars than expected by chance ($p < 1 \times 10^{-20}$; Figure 5B). These results demonstrate that bridge hyperedges are key integrative components in the core-periphery architecture of co-modulated functional connections in the human brain. Also, see Suppl. Figure. S2 for the spatial distribution of bridge hyperedges.

**Developmental effects are concentrated in cluster hyperedges**

Having defined the architecture, anatomy, and topological role of each hyperedge archetype, we next examined whether these co-modulated structures were specifically driven by development-related changes in brain connectivity. To address this question, we tested whether different hyperedge archetypes exhibited differential age-related differences in strength over the period of adolescence. Specifically, we computed the average edge weight within each hyperedge and measured the correlation between edge weight and age, while covarying for in-scanner motion. We observed that correlations with age differed by hyperedge class, with clusters displaying the largest increases in strength over age (1-way ANOVA: $F = 6.56$, $df = 2$, $p = 0.0016$). Post-hoc comparisons with permutation tests confirmed that stars and bridges displayed weaker correlations with age than clusters ($p = 0.0019$ and $p = 0.0381$, respectively). When we examined each cluster separately, we observed that two clusters displayed average strength significantly correlated with age (Figure 4): cluster 2 in the visual system, where the F-statistic *versus* constant model gave $F = 6.34$, $p = 0.00185$, and cluster 5 in the default mode with $F = 17.8$, $p = 2.69 \times 10^{-8}$.

## Discussion

We developed and applied a novel approach to examine high-resolution edge-based hypergraphs in the developing human brain. Hypergraphs are composed of hyperedges, which are groups of functional connections whose strengths co-vary with one another across subjects. We applied this approach to resting state data acquired from 780 youth, uncovering an edge-based core-periphery structure whereby peripheral stars are linked to clusters in the functional core via topological bridges. Stars were centered on specific brain regions and clusters recapitulated well-known cognitive systems including visual, default mode, salience, and cingulo-opercular systems. While stars and bridges could be explained by individual differences across the subject cohort, clusters in the topological core of the hypergraph were driven by development-specific changes in resting state brain dynamics. By treating a functional connection as the fundamental unit of interest, these findings suggest a new conceptualization of brain organization that is not offered by typical network analyses.

**Clusters recapitulate specific functional brain modules**

Cluster hyperedges corresponded to some of the well-known cognitive systems described in the neuroimaging literature, including the visual, motor, default, and cingulo-opercular/salience systems. The default mode network was split into two components, both of which included connections with the central regions of the ventromedial prefrontal cortex and posterior cingulate / precuneus. Notably, one default mode hyperedge was preferentially focused around the medial temporal lobe, paralleling prior accounts of default-mode subsystems (Andrews-Hanna et al. 2010).

Functional HypergraphFunctional Hypergraph

Beyond their anatomical specificity, clusters are also topologically poised to perform specific functions. Indeed, clusters are enriched for highly-connected hub edges (Suppl. Figure S1), and therefore form a relatively stable basis around which all other functional associations in the brain can evolve over development. Such edge cores are conceptually similar to core regions in the brain's "rich club" (Van Den Heuvel and Sporns 2011), and may similarly mediate information transfer over integrating connections involving sensorimotor processes and the default mode network. Clusters are thus well-situated to serve as the backbone of developing information processing capabilities throughout adolescence.

**Stars: local motifs distributed across the brain**

While cluster hyperedges align well with a few of the known large-scale cognitive systems from previous regionally-based network analyses, our high-resolution edge-based network analysis additionally uncovered a novel sub-network type that we refer to as the star hyperedge. Stars are centered around a small number of brain regions (one, two, or three), with edges that radiate outwards. Notably, such star-shaped systems cannot be detected by network analyses where nodes are represented by regions rather than edges: community detection techniques by definition will group together regions with similar patterns of connectivity (Bassett, Wymbs, et al. 2011). Indeed, when community detection analyses of regionally-based networks identify a sub-network with a single node, such a result is generally considered a fragment and not considered further (Bassett, Porter, et al. 2013).

Intuitively, star-shaped sub-networks may represent key partitions of connections that integrate processes from diverse sources within a single node or broadcast to other nodes. Such a regionally-focused account of specialized functional networks is consistent with lesion-based data from both animals and humans, where localized injury may have highly specific functional consequences (Langlois et al. 2006; Alstott et al. 2009). In contrast to the densely-connected core where cluster hyperedges are concentrated, stars are present in the hypergraph periphery. These peripheral stars most commonly occur in regions important for higher level cognitive processes (Van Den Heuvel and Sporns 2011), including the frontoparietal system and other multimodal (Sepulcre et al. 2012) regions across the brain.

**Bridges link stars to clusters**

In addition to star-shaped formations, the hypergraph approach identified the presence of bridge hyperedges for the first time. In contrast to clusters, which have locally dense connections, bridges exclusively link two disparate sets of brain regions.  We found that bridges preferentially linked clusters in the brain's functional core to stars in the periphery. The linking architecture of a bridge hyperedge implies a critical role within the brain's core-periphery framework, potentially facilitating information flow between highly segregated functional systems and regions where distributed higher-order processing occurs.  As with the stars, bridges cannot be identified using typical regional-based network analysis. It is important to note that we did not predict the discovery of these bridges.  However, bridge-like (or bipartite) formations are a frequent feature of other types of systems, including microbial complexity networks (Corel et al. 2016) and microbiome data (Sedlar et al. 2016).

**Development drives co-modulated functional connections**

As a final step, we examined how hypergraphs developed during youth. Developmental differences were concentrated among cluster hyperedges. Specifically, we found significant associations with age in the strength of the default mode and visual hyperedges. These results accord with a pattern of network segregation: the visual and default mode systems have very strong within-network connectivity, and relatively limited connectivity between other brain networks (Gu et al. 2015). Their strengthening during development leads to network segregation that could support a greater diversity of the brain's dynamic repertoire (Betzel et al. 2016) as well as an enhanced capability for adaptation (Mattar et al. 2016).

**Methodological considerations**

Some limitations apply to the current study. While the parcellation was selected for its robustness and prominence in the literature (Power et al. 2011), other schemes are available and may offer additional insights (Bassett, Brown, et al. 2011). In particular, this atlas undersamples subcortical and cerebellar regions, which may be particularly important in development. In addition, while the edge-by-edge hypergraph representation is applicable to all estimates of functional connectivity (Bassett et al. 2014), we employed a pairwise coherence between region time series (Zhang et al. 2016). Finally, inference regarding developmental effects are limited by the use of a cross-sectional dataset; longitudinal research designs would be a useful complement to corroborate the findings reported here.

**Acknowledgement**

DSB acknowledges support from the John D. and Catherine T. MacArthur Foundation, the Alfred P. Sloan Foundation, the Army Research Laboratory and the Army Research Office through contract numbers W911NF-10-2-0022 and W911NF-14-1-0679, the National Institute of Health (2-R01-DC-009209-11, 1R01HD086888-01, R01-MH107235, R01-MH107703, R01MH109520, and R21-M MH-106799), the Office of Naval Research, and the National Science Foundation (BCS-1441502, CAREER PHY-1554488, and BCS-1631550). TDS acknowledges support from the National Institute of Mental Health: R01MH107703 (TDS), R21MH106799 (DSB \& TDS) and Penn Institute of Translational Medicine and Therapeutics (DSB & TDS). The PNC was funded through NIMH RC2 grants MH089983 and MH089924 (REG). The content is solely the responsibility of the authors and does not necessarily represent the official views of any of the
funding agencies.

**Author contributions statement**

T.D.S. and D.S.B. designed research; S.G., M.Z., T.D.S., and D.S.B. performed research; S.G. and D.S.B. contributed new reagents/analytic tools; S.G., M.Y., T.D.S., and D.S.B. analyzed data; and S.G., M.Z, T.D.S., J.D.M., R.E.G., R.C.G., and D.S.B. wrote the paper.

**Additional information**

The authors declare no conflict of interest here.

Functional Hypergraph

# Figures

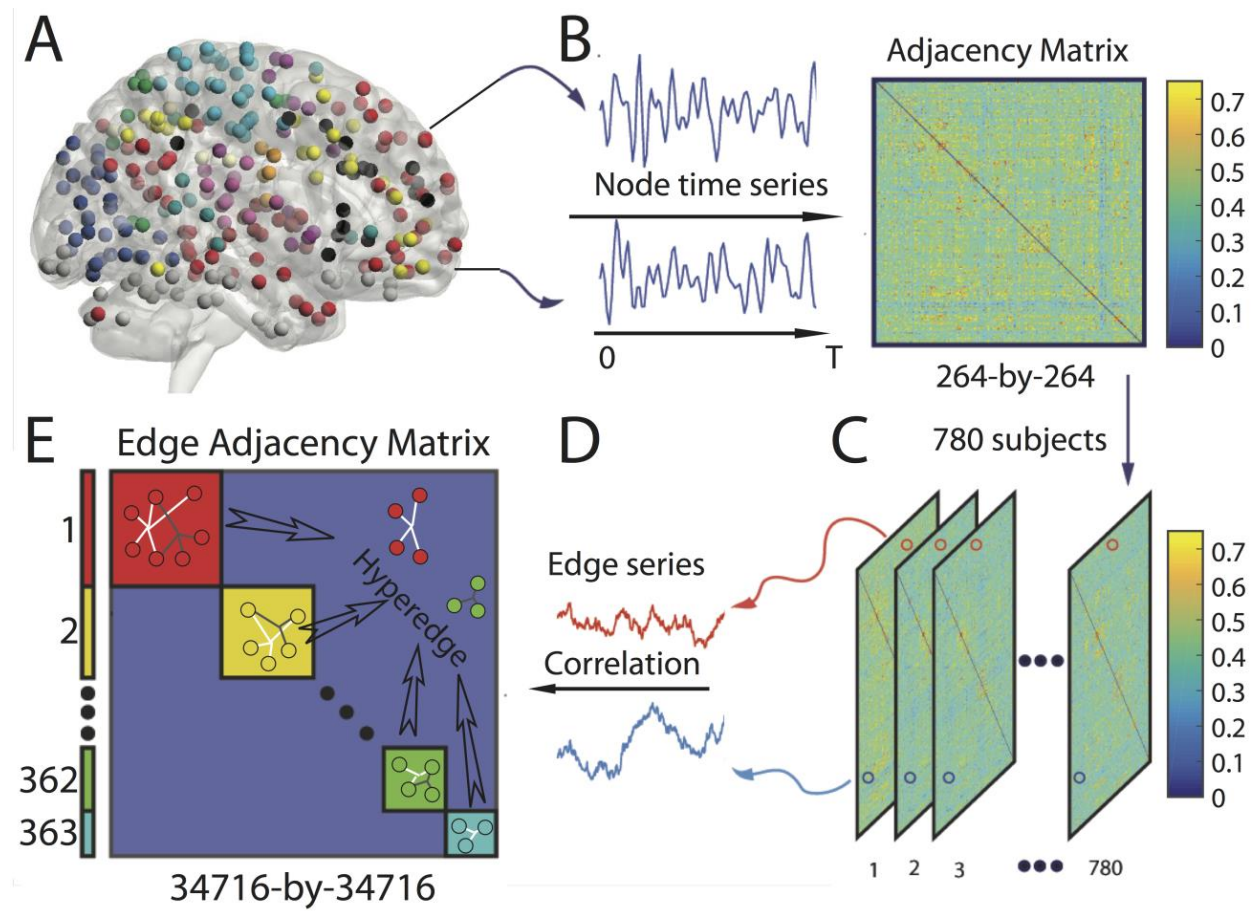

**Figure 1 Schematic of Hypergraph Construction.** *(A)* We first extract time series for each region of interest. *(B)* Next, we calculate the functional connectivity between pairs of regions using a wavelet-based coherence, yielding an adjacency matrix. We perform the steps outlined in panels *(A)* and *(B)* for each of the 780 youth in the Philadelphia Neurodevelopmental Cohort and *(C)* stacked the adjacency matrices across subjects. *(D)* We extract the time series of weights for each edge over subjects. *(E)* Finally, we generate an edge-by-edge adjacency matrix (or hypergraph (Bassett et al. 2014)) by computing the Pearson correlation coefficient between pairs of edge weight time series. A hyperedge is then defined as a cluster of edges that co-evolve with

Functional Hypergraph

one another; we can identify these clusters by applying community detection techniques to the hypergraph.

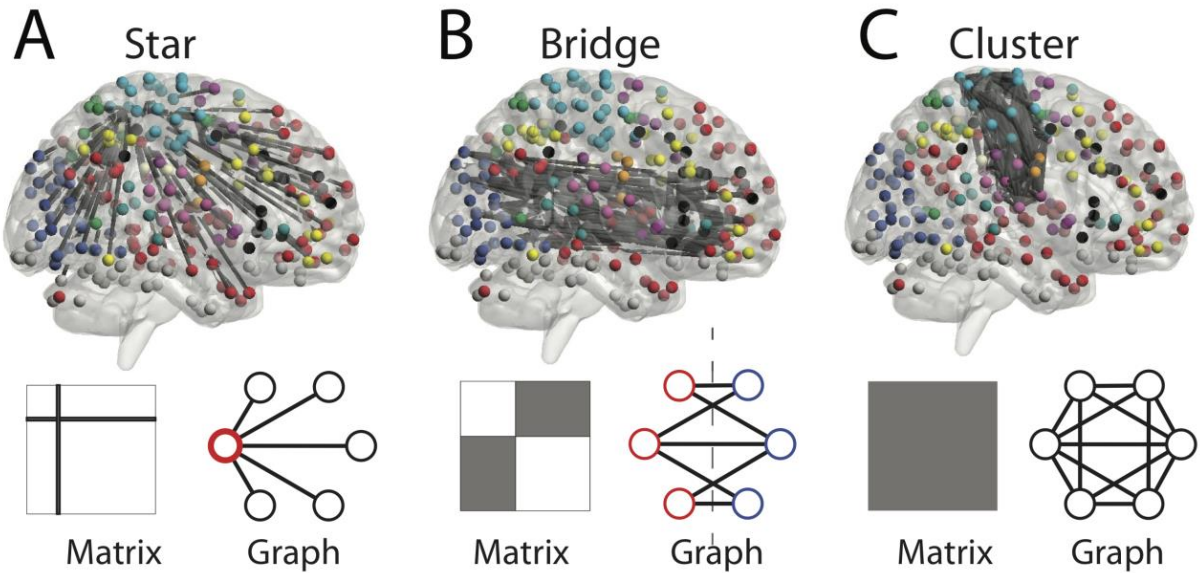

**Figure 2 Co-evolution Archetypes in Functional Connectivity.** Example star *(Left)*, bridge *(Middle)*, and cluster *(Right)* hyperedges with accompanying graph representations.

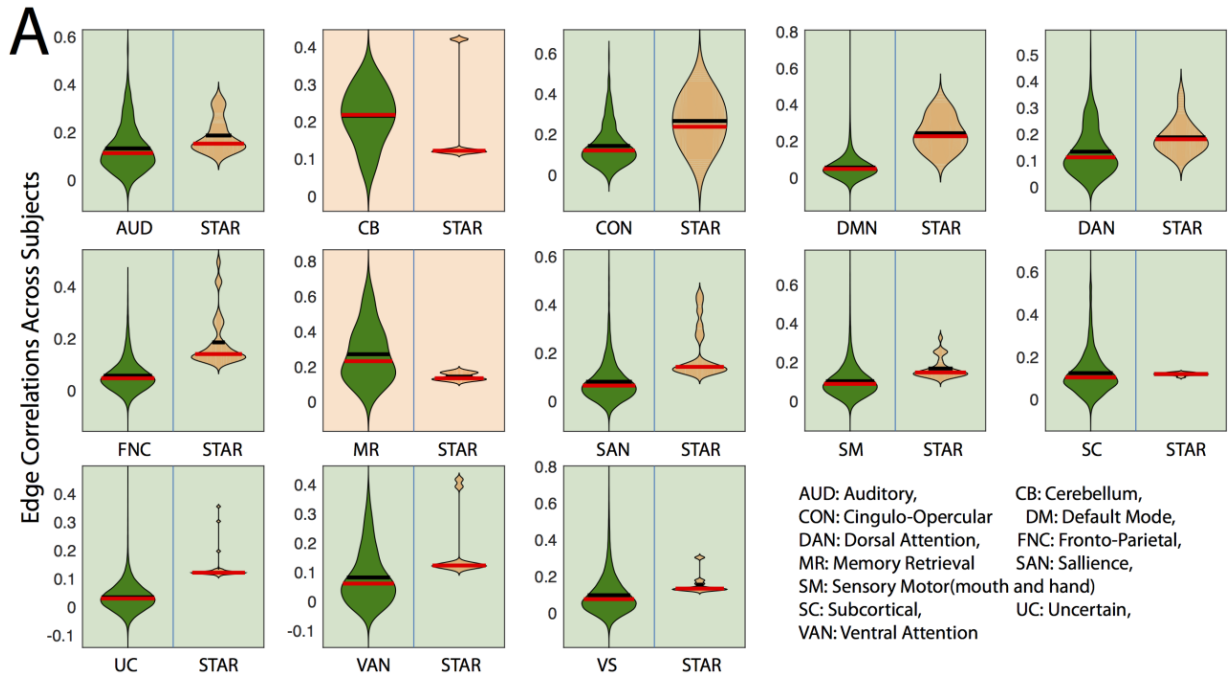

**Figure 3 Edge Correlations Across Subjects.** To investigate whether the star hyperedges displayed higher cohesiveness among edges than well-known cognitive systems or network modules, we compared the star hyperedges and 13 predefined modules (Power et al. 2011) with the null hypothesis that the pairwise similarity of edges in star hyperedges was no higher than that of the edges within the modules. Except for the two smallest systems (the cerebellum and memory retrieval), all other null hypotheses were rejected with the FDR-corrected $q < 0.001$. Results emphasize the higher level of coherence present in star hyperedges than in traditional node-based network modules.

Functional Hypergraph

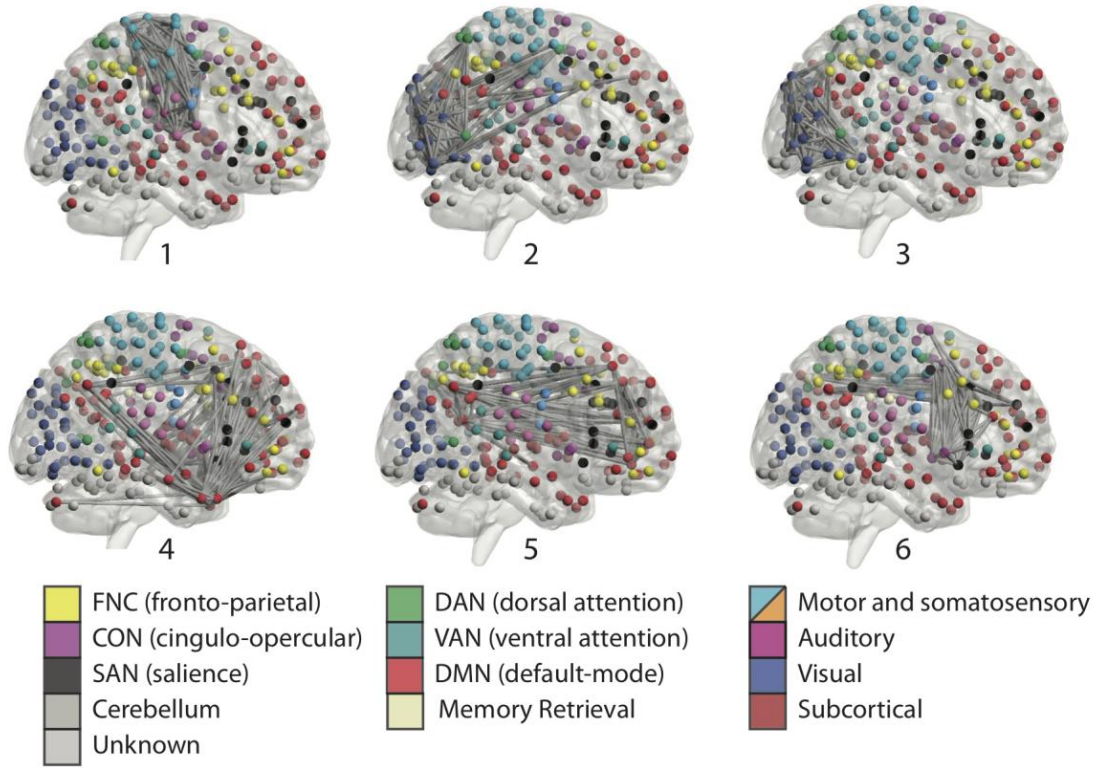

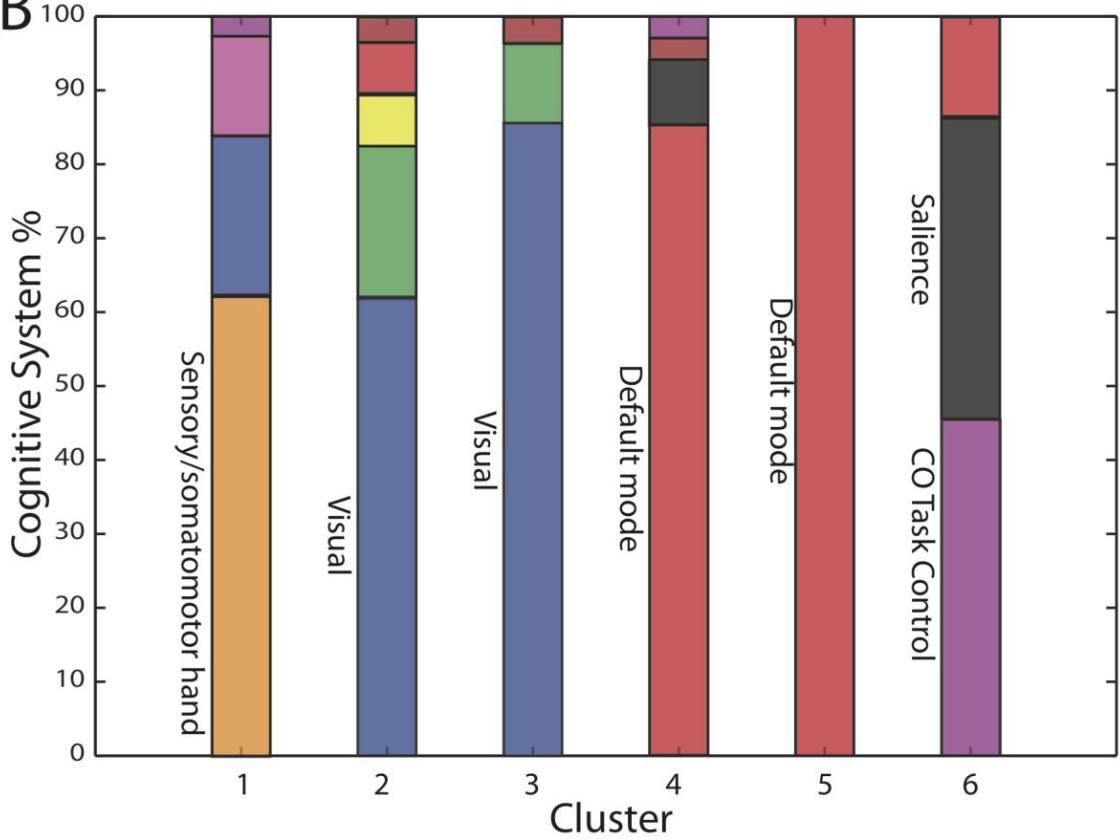

**Figure 4 Anatomical Location of Clusters.** *(A)* Six of the 363 hyperedges were clusters. Each cluster displayed a distinct spatial organization that recapitulated well-known functional systems. *(B)* Each cluster hyperedge connects a set of nodes, and each node (brain region) belongs to a previously-defined cognitive system. Of the nodes present in a cluster, we show the percent that are a part of each cognitive system.

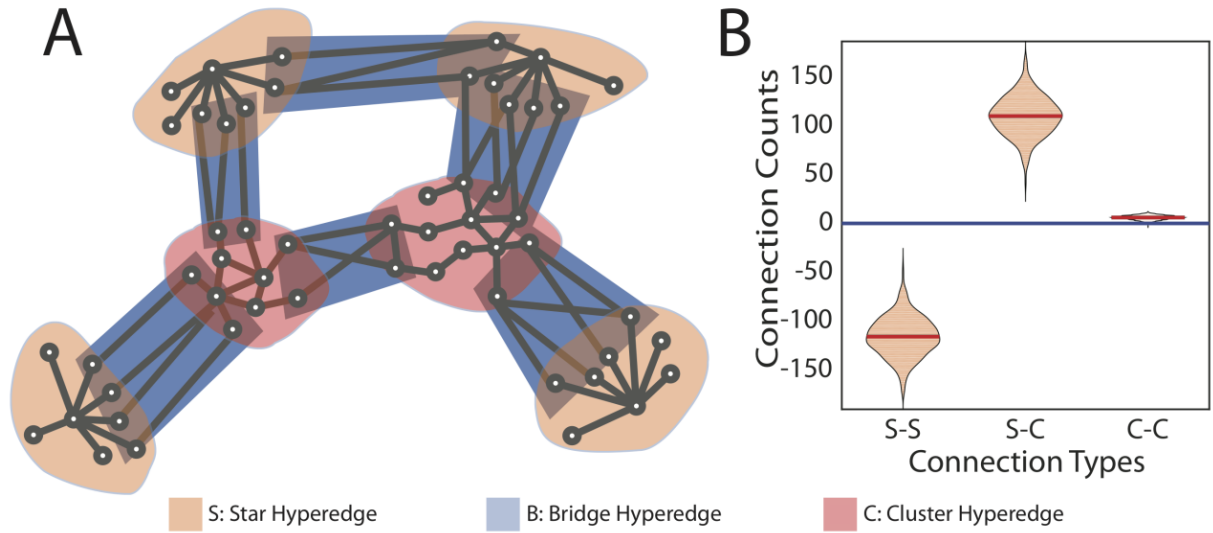

**Figure 5 Bridges Connect Stars to Clusters.** *(A)* A schematic figure shows bridge hyperedges (blue) connect stars (peach) and clusters (pink). *(B)* Testing the intersection of hyperedges versus the null distribution where the nodal occupation is uniformly sampled from all brain regions (see Methods), we demonstrate that bridges are more likely to connect stars with clusters than expected ($p < 1 \times 10^{-20}$). Moreover, bridges are less likely to connect stars to other stars ($p < 1 \times 10^{-20}$).

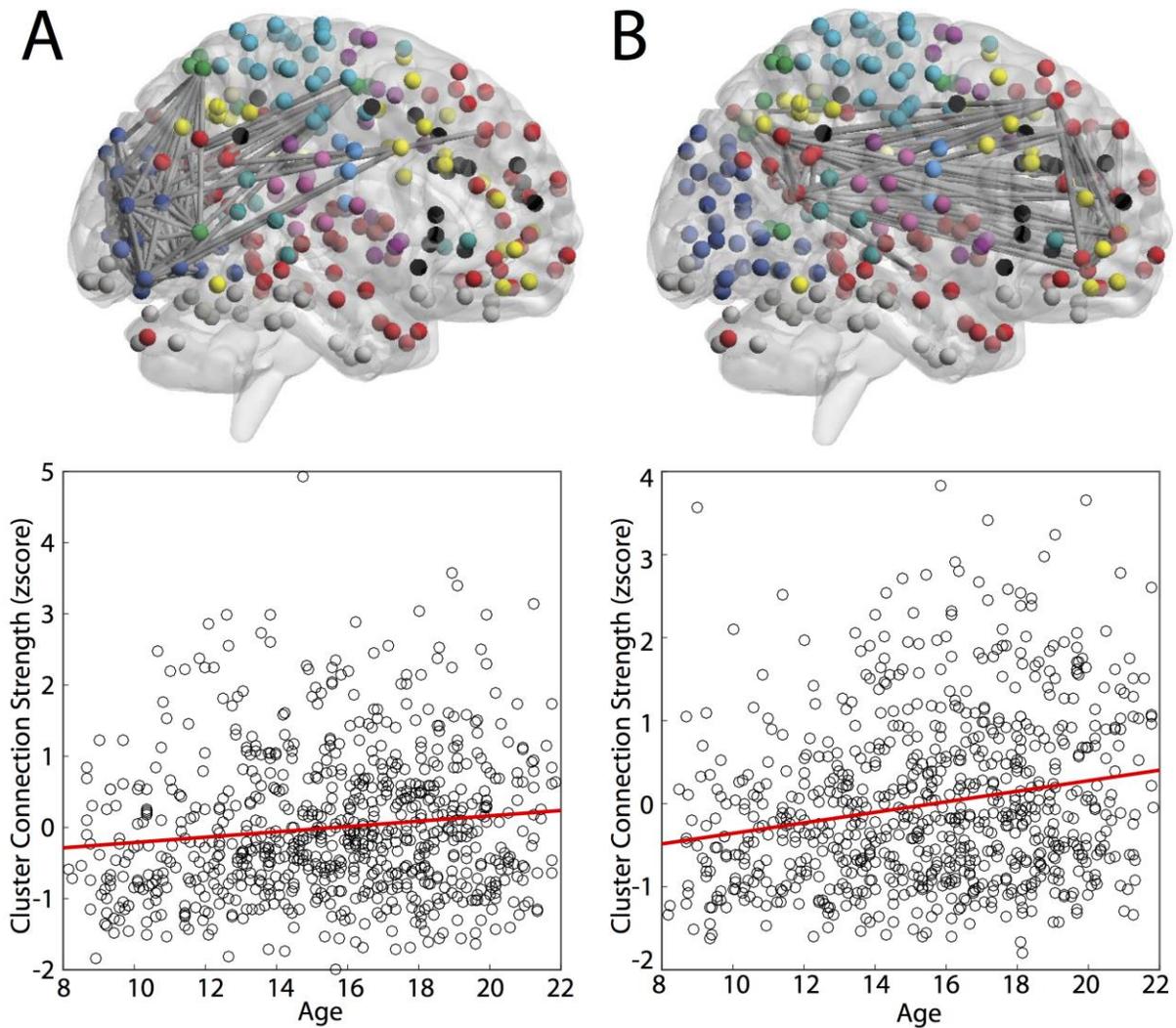

**Figure 6 Cluster Hyperedges Track Developmentally-Driven Co-Modulation of Functional Connections.** Two clusters displayed age-related increased in average strength, as tested by a linear model with age and movement. Panel *(A)* displays the relationship between edge strength and age in cluster 2, composed predominantly of regions in the visual system: $F = 6.34, p = 0.00185$. Panel *(B)* displays the relationship between edge strength and age in cluster 5, composed predominantly of regions in the default mode: $F = 17.8, p = 2.69 \times 10^{-8}$.